# Geometric Deep Learning for Molecular Crystal Structure Prediction

*Michael Kilgour*[(1)], *Jutta Rogal*[(1)(2)], *Mark Tuckerman*[(1)(3)(4)(5)]∗

[(1)] *Department of Chemistry, New York University, New York, NY 10003, USA*
[(2)] *Fachbereich Physik, Freie Universität Berlin, 14195 Berlin, Germany*
[(3)] *Courant Institute of Mathematical Sciences, New York University, New York, NY 10012, United States*
[(4)] *NYU-ECNU Center for Computational Chemistry at NYU Shanghai, 3663 Zhongshan Rd. North, Shanghai 200062, China*
[(5)] *Simons Center for Computational Physical Chemistry at New York University, New York, NY 10003, USA*

**Abstract**

We develop and test new machine learning strategies for accelerating molecular crystal structure ranking and crystal property prediction using tools from geometric deep learning on molecular graphs. Leveraging developments in graph-based learning and the availability of large molecular crystal datasets, we train models for density prediction and stability ranking which are accurate, fast to evaluate, and applicable to molecules of widely varying size and composition. Our density prediction model, MolXtalNet-D, achieves state of the art performance, with lower than 2% mean absolute error on a large and diverse test dataset. Our crystal ranking tool, MolXtalNet-S, correctly discriminates experimental samples from synthetically generated fakes and is further validated through analysis of the submissions to the Cambridge Structural Database Blind Tests 5 and 6. Our new tools are computationally cheap and flexible enough to be deployed within an existing crystal structure prediction pipeline both to reduce the search space and score/filter crystal candidates.

# 1. Introduction

The properties of molecular crystals, including physical and bio-active features, depend sensitively on the details of the crystal structure[1]. In order to ensure safety and efficacy of drugs or engineer the desired properties into functional organic materials, such as organic semiconductors, it is necessary to identify the stable polymorphs into which a given molecule may crystallize before deployment. Due to the large number of plausible ways for atoms and molecules to pack together, it is not straightforward, in general, to predict crystal structures from only single-molecule information.

Crystal structure prediction (CSP) generally must tackle two problems: (i) searching for likely structures using random or grid sampling, Markov chain Monte Carlo (MCMC)[2] or genetic algorithms (GA)[3–8], and (ii) scoring or ranking found structures from energetics obtained via empirical force-fields or quantum chemistry (QC). Force field potentials may be fast to evaluate but they are generally lacking in either accuracy and/or general applicability. QC calculations have higher accuracy and general applicability but are rather costly to run for large numbers of proposed structures. The search needs to explore many degrees of freedom: the number of molecules in the asymmetric unit, $Z'$, the position, orientation, and conformation of the molecule, the size and shape of the unit cell, and the space group. The high dimension of the search space, combined with the cost of accurate ranking calculations renders CSP an expensive proposition, with one group in the recent Cambridge Structural Database Blind Test 6[9] expending 30 million CPU hours on a single molecular target[10]. There is a clear opening for accurate methods supporting this search that are applicable to a variety of systems, inexpensive to employ, and help reduce the computational cost.

Standard approaches to the ranking problem fall into two broad categories: Energy-based approaches compute the total energy of a structure as a function of the atomic coordinates employing, for example, general purpose force-fields such as the GAFF[11], purpose fit molecule-specific force-fields[12], or ab initio tools such as density functional theory (DFT). Purely structure-based approaches (also called "topological" or "geometric"), on the other hand, generate a score directly from the atomic coordinates but without requiring an energy evaluation. Our approach follows this second path, obviating the need for an energy evaluation and using the power of modern statistical techniques to enable potentially high accuracy at low cost.

Protocols of geometric analysis can be understood based on increasing orders of structural correlation functions. At the lowest level is the 1-body spatial correlation, yielding the average density. If a sample is obviously outside the range of non-porous molecular crystal densities (packing coefficient $c_{pack}$ roughly $0.55 < c_{pack} < 0.85$), it is rejected as implausible. Stepping up one level to pairwise correlations, samples are rejected if any pair of atoms is significantly closer than the sum of their respective van der Waals radii. In a set of random crystal packings, adherence to both of the above conditions is rare. Since these conditions are computationally inexpensive to check, they make a highly efficient coarse-grained filter in molecular crystal search. Although these notions are generally accepted in the field of molecular crystals, they are nevertheless important as a foundation for our discussion of higher order approaches and a preview of what can be done with purely structural information.

Several structure-based methods have been developed going back to at least 1998[13–17]. The common themes are (i) the exploitation of the large and growing availability of experimental molecular crystal structures, most importantly those available in the Cambridge Structural Database (CSD)[18], and (ii) the modelling of important intermolecular distances, usually via some pairwise or two-body correlation function such as the spatial distribution, radial distribution, or fingerprinting. These approaches leverage the observation that intermolecular distances are similar between atoms in similar environments. Concretely, one can extract the distribution of pairwise interatomic distances from experimental structures and then score proposed structures or predict their likelihood according to this distribution.

The two key limitations common in prior structural approaches are (i) the problem of explicitly enumerating all the pairs of atom types (elements, functional groups, fragments, or environments) over which to model, and (ii) use of low-order structural correlations, generally limited to pairwise distances.

The first issue arises due to the combinatorial explosion of atom-pair combinations, together with the concomitant decrease in available training examples as one considers more and more specific atom types. In the simplest case, all atoms are considered to be the same type, yielding a single all-to-all correlation function. This does not accurately represent real materials, as for example, C-C and C-O distances have significantly different distributions. Going further, each element could be assigned a separate atom type, resulting in $\frac{1}{2}(N_{z_a}^2 + N_{z_a})$ unique radial distribution functions, with $N_{z_a}$ being the number of elements considered. This still can be improved, as for example methyl and aromatic carbons have different distributions of interatomic distances. One can continue in this manner, using more nuanced atom type definitions, while ballooning the number of distinct atom pairs. Ultimately, it is impossible to fully capture the smooth and continuous variation in local atomic environments by such a discrete system.

On the second issue, consideration of strictly low-order correlations results in a model that cannot capture important physical features such as bond angles, directional bonding, and many-body correlations in general. Similar to the increasing complexity of atom types, this could be addressed by explicitly modelling higher order correlations such as bond and dihedral angles (3- and 4-body), but this considerably increases model complexity, and, due to the curse of dimensionality, sparsifies the space of training data.

Building a simple empirical model over high order correlations with highly specific atom types is technically possible but, due to data sparsity, unlikely to result in a robust and general model. Modern deep learning approaches on the other hand provide the tools to address these weaknesses, capturing atomic environments in a continuous representation and learning high-order correlations, while generalizing well to unseen data. In particular, we deploy advances from the fields of geometric deep learning and learning on graphs using deep graph neural networks (DGNNs).

The purpose of a graph neural network is to learn some function of a graph in the space of its nodes (vertices), which are connected by edges according to some structural logic. As an example, in a sentence, each word could be assigned as a node, and edges as all-to-all semantic connections between them. In the simplest molecular graphs, atoms are embedded as nodes and covalent bonds between atoms as edges.

We follow in the lineage of DGNN models such as SchNet and PhysNet[19,20], encoding each atom in a molecule or molecular crystal as a node in a graph, assigning directional edges between atoms if they are within a cutoff range, $r_c$, and featurizing these edges with a spatial embedding function, discussed in Section 2. Nodes repeatedly pass messages between each other in synchronized steps called graph convolutions. These messages carry information from node to node, conditioned on their edge embeddings, which incorporate information on the relative 3D positions of atoms. In this way, atoms aggregate information about their local atomic environments, including the specifics of the molecular geometry.

After a series of message passing steps, the model aggregates information from all the nodes to a single vector representing the whole graph. A feedforward neural network can then learn a desired function based on the graph readout. If supplied with sufficiently rich features, sufficiently large DGNNs can learn arbitrarily complex functions in the space of 3D point clouds up to the geometric limitations of the chosen architecture[21–24]. By construction, our models are invariant to permutations in atom ordering and global translations, rotations, and inversions; they are, therefore, suitable for learning scalar functions on molecular graphs. There is a growing literature of so-called equivariant DGNNs[25–28] for learning vector functions such as forces on molecular graphs, but we see no immediate need for this extra capability for our applications.

DGNNs have been effectively used in the past for a wide variety of learning tasks on molecules and atomic crystals including property prediction, stability evaluation, structure generation, and more (see Ref. [23] for a very recent review). In this work, we demonstrate how geometric deep learning techniques may be extended to molecular crystals by modelling two crucially important crystal properties. Based on the previous discussion, we develop a new molecular crystal DGNN model for crystal ranking, MolXtalNet-S, which inputs the proposed atomistically detailed unit cell structure and returns a stability score. We also undertake a more traditional molecular bulk property prediction, training a separate DGNN model, MolXtalNet-D to predict the density of a given molecular crystal, given only the molecule conformation. Prior works have also used machine learning methods to predict the density of a crystal based on molecule information[29–31]. In these approaches, models were provided only whole-molecule features, such as the presence of certain fragments and the molecule surface area, as opposed to learning a geometric representation directly from atom positions.

Our two models for density prediction and stability scoring could greatly accelerate molecular CSP by radically reducing the parameter search space via a constraint on the range of likely densities and by providing a tool for ranking proposed crystal structures. Both models are cheap to evaluate and generally applicable to a wide range of systems, including large and small molecules with light and heavy atoms. The crystal scoring model, MolXtalNet-S, accepts crystals in any space group with one molecule in the asymmetric unit, $Z' = 1$.

This paper is organized as follows. In Section 2, we introduce the DGNN models we employ for molecular crystal modelling. In Section 3, we explain in detail how we construct the dataset. In Sections 4 and 5, we show how our DGNNs perform on density prediction and crystal scoring tasks, respectively, evaluating samples from the CSD and from the CSD Blind Tests 5 and 6. We conclude in Section 6.

## 2. Models and Methods

### 2.1 *Molecule Graph Model*

Our graph models follow the general framework of SchNet[19] and related models, wherein atoms are encoded as nodes with directed edges featurized by some embedding function determined by the local geometry. We have made several customizations, mostly to improve expressiveness and flexibility of the model, which we will elucidate here.

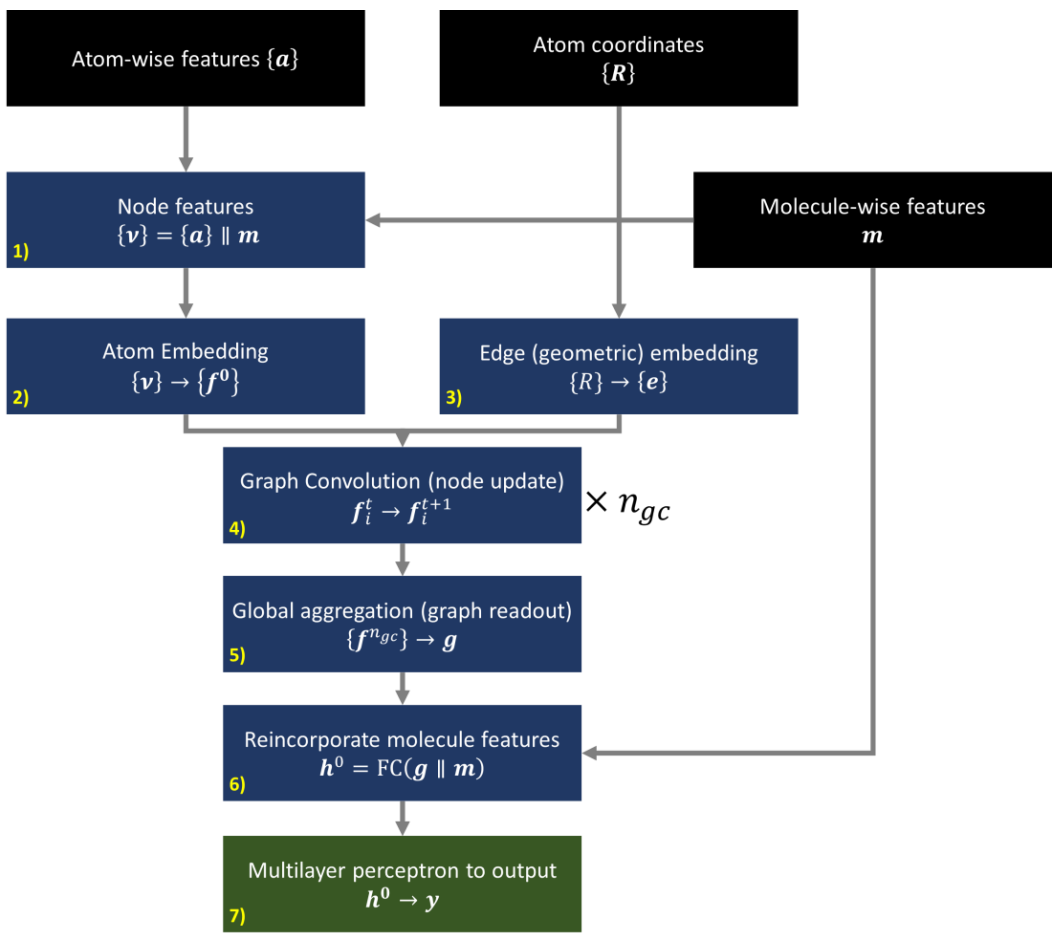

**Figure 1:** Schematic of our MolXtalNet graph neural network architecture. $n_{gc}$ is the number of graph convolution layers in the model.

The inputs to the model are the set of each atom's feature vectors, $\{a\}$, each with length 8, the set of atom coordinates, $\{R\}$, and a feature vector for molecule-scale features, $m$, with length 16 (see full feature lists in Appendix A). Molecule features are concatenated to all atoms, producing $\{v\}$, which gives each node initial global context. Atomic numbers (the first dimension in $v$) are then replaced by a vector embedding, which is re-concatenated back to the node feature vector and subsequently fed through a fully connected layer and activation,

$$\begin{aligned}\boldsymbol{\mu}_i &= \text{EMB}(v_i^0) \parallel \boldsymbol{v}_i^{j>0}, \\ \boldsymbol{f}_i &= \sigma(\text{FC}(\boldsymbol{\mu}_i)),\end{aligned} \qquad (1)$$

with □ ∥ □ as vector concatenation, A fully connected linear layer is defined as

$$\text{FC}(x) = W \cdot x + b, \tag{2}$$

with $x$ an input vector and $W$ and $b$, respectively, a learnable weight matrix and bias vector; the activation function, $\sigma(x)$, is taken as the leaky ReLU throughout. Except for $W$, all boldface variables are one-dimensional vectors.

There are several approaches to edge embedding, wherein geometric information is introduced to the model. We tested the hierarchy of embedding functions from SchNet, to DimeNet, and SphereNet, incorporating increasingly higher-order structural correlations in the edge embeddings[19,20,32–34]. Ultimately, the most consistent and stable model included 2-body (radial) embeddings only, using the Bessel basis formulation from DimeNet[32] with 32 basis functions, omitting explicit angular information. For both tasks, we use a model with four graph convolution layers and four fully connected layers. The feature depth throughout (vectors $f, g, h$) is 256, except within the message passing step where it bottlenecks to 128. Layer normalization, $N(x)$, and dropout, $D(x)$, of 0.1 are used both in the graph convolutions and the output multilayer perceptron (MLP).

In the "graph convolution" or "message passing" stage, the model combines geometric information of the molecule with the features provided in the nodes. The node information is first bottlenecked from 256 dimensions to the message size of 128 by the node-to-message FC layer, $\text{FC}_{n \to m}$, and the radial embedding $e_{ij}$ between the source node indexed by $j$ and target node indexed by $i$ has its dimension boosted from 32 to 128 in the same manner by the edge-to-message FC layer, $\text{FC}_{e \to m}$,

$$F_i^t = N\left(\text{FC}_{n \to m}(f_i^t)\right), \tag{3}$$

$$E_{ij}^t = \text{FC}_{e \to m}(e_{ij}), \tag{4}$$

with $F_i^t, E_{ij}^t$, the resulting 128-dimensional vectors representing nodes and edges respectively. These are subsequently used to construct messages in the $t$th graph convolution,

$$f_i^{t+1} = f_i^t + \sum_{j \in r_c} \text{GCONV}(F_i^t, F_j^t, E_{ij}^t), \tag{5}$$

$$\text{GCONV}(F_i^t, F_j^t, E_{ij}) = \text{FC}_2\left(\sigma\left(N\left(\text{FC}_1(F_i^t \parallel F_j^t \parallel E_{ij}^t)\right)\right)\right), \tag{6}$$

With $\text{FC}_1, \text{FC}_2$, as two FC layers with input:output dimensions of 384:128 and 128:256 respectively, and $j$ the index for message sources running over the nodes within $r_c$, the convolution radius about node $i$, throughout taken as 6Å.

After $n_{gc}$ graph convolutions, we aggregate all nodes of the graph to a single 256-dimensional feature vector. Following recent work on problems of expressivity of global aggregation[35,36], we employ a combination of global aggregators, max, sum, mean, and self-attention (SA), in parallel, concatenating the results and passing through a fully connected layer,

$$\text{SA}_k\{f\} = \sum_i^{\text{nodes}} \text{softmax}\left(\text{FC}_2(\sigma(\text{FC}_1(f_i)))\right) \odot (\text{FC}_3(f_i)), \tag{7}$$

$$g = \text{FC}(\text{MAX}_k\{f\} \parallel \text{SUM}_k\{f\} \parallel \text{MEAN}_k\{f\} \parallel \text{SA}_k\{f\}) \tag{8}$$

with the aggregation operations running over the feature depth index $k$.

The resulting feature vector, $g$, encodes what the model has learned from the molecule or molecular crystal graph. We recombine this vector with the original molecule-level features and pass the result through an MLP with $n_{gc} + 1$ layers returning the final output,

$$h^0 = \text{FC}(g \parallel m), \tag{9}$$

$$y = \text{MLP}(h^0),$$

$$h^{t+1} = h^t + \text{FC}_2\left(D\left(\sigma\left(\text{FC}_1(N(h^t))\right)\right)\right), \tag{10}$$

$$y = \text{FC}(h^{n_{Gc}}),$$

where $y$ is either the 1-dimensional regression output for bulk density estimation in MolXtalNet-D or the 2-dimensional output representing the probability a given molecular crystal is "real" or "fake" in MolXtalNet-S.

*2.2 Molecular Crystal Graph Model*

There is an open question of how to encode the symmetry properties and periodicity of molecular crystals most efficiently in a GNN, which, to our knowledge, has not been previously addressed. In the study of atomic crystals, there are methods for crystal graph construction,[37–40] which compactly represent the full periodic structure by a subset of atoms with self-connections representing interactions between periodic images. One could consider doing the same for molecular crystals; however, the interatomic distances between periodic images in molecular crystals are often rather long compared to typical graph convolution ranges ($r_c \approx 5 - 10$ Å), and extending this range introduces potential issues, as the convolution window includes more and more atoms, proportional to $r_c^3$.

We, therefore, develop a molecular crystal graph convolution (MCGC) taking inspiration from padding techniques in image processing (see diagram in Figure 2). We first identify and separately label each atom within a *N*x*N*x*N* (in practice, generally 3x3x3) supercell as follows: 0 for atoms in the molecule within a chosen asymmetric unit, which we call the "canonical conformer", 1 for atoms within $r_{max} + r_c$ from canonical conformer centroid, with $r_{max}$ as the maximum distance between the centroid and any atom in the molecule, and 2 for atoms outside this range. The crystal graph is then constructed of nodes labelled 0 or 1, with 2s discarded. Directional edges are allowed between all atoms labelled 0, and from the outside-in, from atoms labelled 1→0. In physical terms, intramolecular messages are allowed as in a standard molecular DGNN, and intermolecular messages are allowed only toward the canonical conformer from its symmetry images. Periodicity is enforced by overwriting the feature nodes of all atoms in the canonical conformer to their symmetry images in the rest of the 3x3x3 supercell after each node update. This exploits the periodicity of the system to achieve the same result as if we had performed a convolution on an infinitely large molecular crystal graph. Through repeated graph convolutions in this manner, we can aggregate structural and chemical information in the usual way for DGNNs out to a range of approximately $r_c \times n_{gc}$, where $n_{gc}$ is the number of graph convolution layers in the model. It is good practice to ensure that the supercell size at least encapsulates all atoms which could be labelled '1' according to the above procedure.

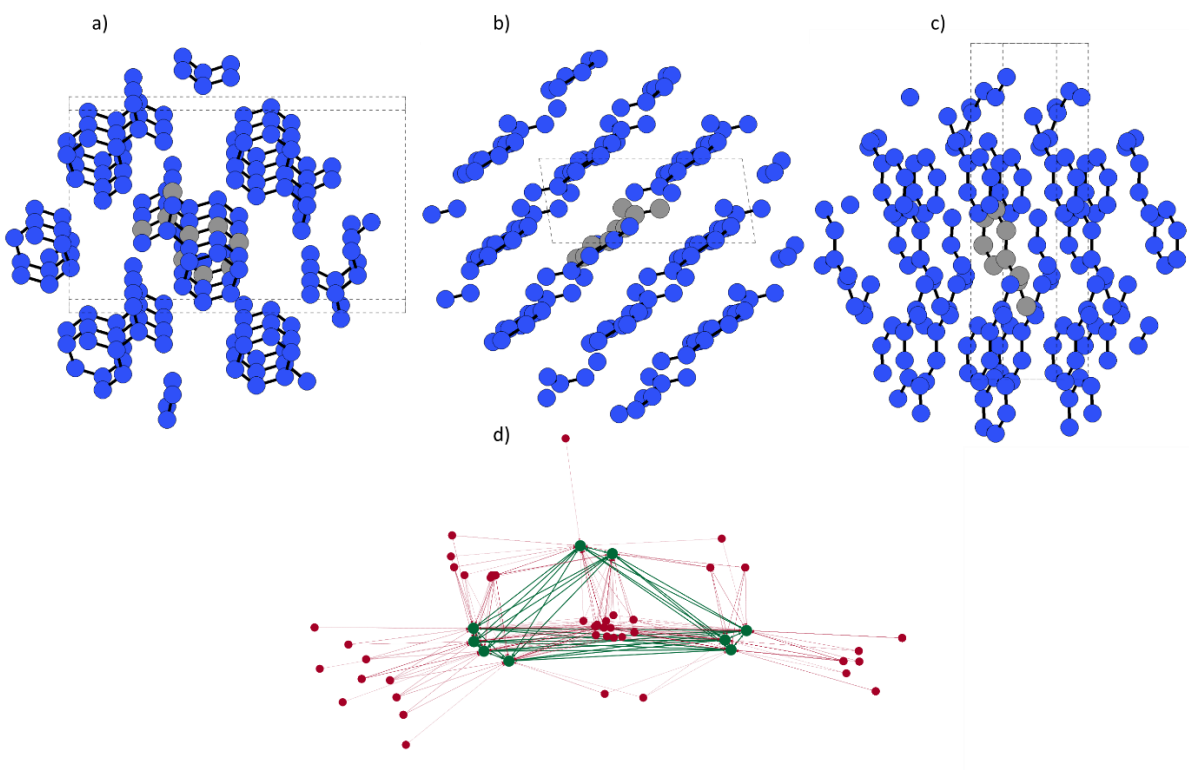

**Figure 2:** Panels a)-c) show views along the reciprocal axes of CSD structure NICOAM03. The grey molecule is the canonical conformer, and blue all the symmetry images that could potentially participate in graph convolution. Messages pass to and within the canonical conformer, and after node update the nodes' feature vectors are copied to the symmetry images.

Panel d) shows the Kamada Kawai visualization of the directed graph, with green nodes and edges as the canonical conformer and intramolecular connections, and red nodes and edges and symmetry images and intermolecular connections.

Allowing for intermolecular outside-in messages only in the final layer of the crystal DGNN and leaving all prior layers as intramolecular only results in a small but consistent improvement in the test set loss. In this setup, the final node vector is computed as the output of the graph convolution, omitting the residual, and the source index, $j$, runs over only nodes outside the canonical conformer,

$$f_i^{n_{gc}} = \sum_{j \in r_c} \text{GCONV}\left(F_i^{n_{gc}-1}, F_j^{n_{gc}-1}, E_{ij}^t\right). \quad (11)$$

The idea behind this choice is that the model will learn first a detailed representation of the molecular environment. As all of our training and evaluation data comprise realistic molecules, the intramolecular information is of limited utility for crystal ranking, except as a context for intermolecular correlations. By separating intramolecular from intermolecular interactions in a hierarchy of structural correlations, we help the model to focus on the intermolecular factors which determine crystal validity. Of course, this is an architectural choice which may vary depending on the problem under study.

*2.3 Crystal Generation*

In addition to positive examples of "real" experimental crystal structures from the CSD, we require negative examples to train a discriminator model. To this end, we developed a molecular crystal parameterization scheme and two generative approaches for synthesizing "fake" molecular crystals. Both approaches work in the space of the 12 molecular crystal parameters, $C$, defined below (see, also, Appendix B). The first approach samples crystal parameters from a multivariate Gaussian distribution fit to the CSD statistics (Gaussian generator). The second approach works by adding a small amount of Gaussian noise to the crystal parameters of existing CSD structures and rebuilding the unit cell (distorted crystal generator). The Gaussian generator produces generally low-quality samples which fail to respect atomic vdW radii, though with reasonable density. The quality of samples from the distorted crystal generator depends on the magnitude of the applied distortion.

The parameters, $(a, b, c)$ and $(\alpha, \beta, \gamma)$ are the standard crystal cell parameters corresponding to the lengths and internal angles of the unit cell vectors. To analyze and generate molecular crystals in a consistent and repeatable manner, we define an additional six parameters to fix the position and orientation of the molecules in the unit cell. For a crystal with $Z' = 1$ and $Z$ molecules in the unit cell, there are $Z$ possible choices for the parameterization. We assign the molecule with the center of geometry closest to the origin in fractional coordinates as the "canonical conformer". This choice allows consistently repeatable crystal generation and editing, although a different parameterization of the canonical asymmetric unit may be more useful in different circumstances. The fractional coordinates $(\bar{x}, \bar{y}, \bar{z})$ designate the center of geometry of the canonical conformer and $(\phi, \psi, \theta)$ are the angles that characterize the orientation of the canonical conformer from a standardized initial orientation. The standardized initial orientation is defined by aligning the principal inertial axes of the molecule with the cartesian axes. In general, this assignment is ambiguous since the direction of the principal inertial axes is arbitrary. To address this, we employ a slightly modified definition of the principal inertial axes that consistently returns vectors with the same relative directions for a given molecule. A vector is drawn from the centroid to the most distant atom and the principal inertial vectors directions are chosen so to have a positive overlap with it. If the overlap of some principal vector is nearly or exactly zero, as in a 2D geometry, the vectors' directions are set via the right-hand rule. If the resulting principal inertial axes are left-handed, the initial position aligns them instead with $(x, y, -z)$, which respects the molecular symmetry.

Since we consider molecular conformers to be rigid bodies in this work and crystals to be perfectly ordered, these twelve parameters, $\boldsymbol{C}$, plus the choice of space group, completely specify the crystal structure. We accordingly built a fast, differentiable, and parallel PyTorch tool both for extracting such parameters from existing crystals and for generating explicit atomistic supercells given these parameters plus the molecular conformation. Details of the cell builder are provided in Appendix B.

The Gaussian generator uses the covariance statistics from the CSD-derived training dataset to fit a 12-dimensional multivariate Gaussian model, with the cell vector lengths substituted with a reduced set $(a, b, c) \rightarrow (a', b', c')$,

$$a' = \frac{a}{V_{mol}^{\frac{1}{3}} \cdot Z^{\frac{1}{3}}}. \tag{12}$$

This makes the sampling invariant to the number of molecules in the unit cell, Z and the molecule volume, greatly improving the average density of proposed crystals.

To enrich the training data with more plausible fakes, we implemented the distorted crystal generator, which applies a tunable amount of distortion to an existing crystal structure. Starting from an experimental crystal structure, we determine the crystal parameters, $\boldsymbol{C}$, and standardize them according to the dataset statistics, $\boldsymbol{C}_{std} = \frac{(\boldsymbol{C} - \overline{\boldsymbol{C}})}{\sigma_{\boldsymbol{C}}^2}$. We then add 12-dimensional Gaussian noise, scaled by a linear factor $\boldsymbol{C}_{std,dis} = \boldsymbol{C}_{std} + \mathcal{N}^{12} \times c_{dis}$, destandardize the parameters, and reconstruct the unit cell in our usual way. For large values of the distortion factor, $c_{dis} \approx 1$, the samples are essentially random, and for very small distortions, $c_{dis} \approx 0.001$, the samples are too similar to the experimental structures and are not useful as negative examples.

3. Data processing

Our data processing pipeline constructs training and evaluation datasets from either the CSD or directly from collections of .cif files. Crystal structures pass through three filtering and featurization steps in preparation for model training, with the relevant filters listed in Tables 1 and 2 below. The first processing step is the assignment of a unique identifier and the collection of crystal features, such as the space group, cell parameters, and the coordinates for a single conformer and a complete unit cell using the CSD Python API. As this work is a proof of concept for our approach, we limit ourselves to crystals with one molecule in the asymmetric unit, $Z' = 1$. Much of what follows would function straightforwardly for $Z' \neq 1$. At the first level, the filtration catches various straightforward errors, as shown in Table 1.

Table 1: Crystal processing filters

| Filter | Condition |
| --- | --- |
| Entry is empty | Not allowed |
| Entry missing atoms | Not allowed |
| $Z'$ | 1 |
| Wrong number of molecules / components in entry | Not allowed |
| Structure is polymeric | Not allowed |
| Entry is missing 3D structure | Not allowed |
| Unit cell generation fails | Not allowed |

The second processing step is analysis and featurization of the molecular conformer itself. Many atom- and molecule-scale features which will be used as training inputs are computed here, mostly using the RDKit Python package[41], including atom electronegativity, molecule volume, molecule inertial moments etc., see Appendix A for details. Two important filters are applied here; first, as RDKit is the main featurization engine, the molecules must be recognized as valid structures by the package. There are several possible reasons why RDKit may reject a molecule, including errors in kekulization or issues with number of covalent bonds per atom.

The second filter applied at this step is the deletion of all hydrogen atoms. Given that hydrogens fill important chemical and structural roles, it is not ideal to exclude them from the training data. However, inconsistency in whether or not hydrogen positions are included in the CSD, uncertainty about the precision of these positions, and the inability to reliably place implicit hydrogens make it difficult to include them in a consistent way. We do however include hydrogen bond donor/acceptor labels in our atom-wise featurization, and the number of hydrogen bond donors and acceptors per molecule in molecule-wise featurization. Even without explicit hydrogens, our results are rather good, and the model can in principle infer the occupied volume and directional hydrogen bonding given the context of surrounding atoms. It is difficult therefore to estimate the performance difference we would observe including precisely placed hydrogens.

The prior two steps are rather slow, generally taking hours to extract on a single CPU for the full CSD and are therefore done before training. Using the above filters, we yield 313,000 featurized structures from an initial set of 1.21M samples. The third and final step is done at runtime and allows us to quickly pare down the dataset into the relevant subset for a particular model run. We will give more details on these choices when discussing results. For all training runs, the dataset is split 80:20 into train and test datasets, and we repeat most experiments over multiple dataset and model seeds to ensure robustness.

Table 2: Filters applied at runtime

| Filter | Condition |
| --- | --- |
| Blind Test 5 and 6 targets | Not allowed |
| Molecule is organic | Allowed |
| Molecule is organometallic | Allowed |
| Molecule max # atoms | 100 |
| Molecule max atomic number | 100 |
| Z value | Z=Wyckoff multiplicity |
| Z value | $0 < Z \leq 18$ |
| Crystal packing coefficient, $c_{pack}$ | $0.55 < c_{pack} < 0.85$ |
| Crystal has disorder | Not allowed |
| Nonstandard space group settings | Not allowed |
| Multiple polymorphs per entry | Not allowed |
| Missing R-factor in entry | Allowed |
| Exactly overlapping atoms | Not allowed |

Our training dataset is built in this way, starting from the CSD. We also construct evaluation datasets using all the publicly available submissions to the CSD Blind Tests 5 and 6[10] which pass the above filters (approx. 26k and 6k structures, respectively). Naturally, this excludes submissions for multicomponent structures such as Target XIX. We pull the Blind Test target structures directly from the CSD and exclude them from our training sets.

The dataset generated via the above procedure contains 160k samples. The dataset is, like the CSD itself, dominated by space groups $P2_1/c$, $P2_12_12_1$, $P-1$, $C2/c$, but this is not a major issue here, as neither of our methods depend on the particular crystal symmetries, only on the arrangement of molecules in space.

## 4. Density Prediction

We use the crystal density prediction model, MolXtalNet-D, described in Section 2 to infer information about the crystal formed by a given molecule using only single-conformer data. This is an appealing approach since prior knowledge about the crystal can dramatically narrow the search space. MolXtalNet-D is able to successfully model the packing coefficient, and thereby the density, of molecular crystals. Despite extensive testing, we have not yet reliably modelled any additional properties relating specifically to crystal symmetry (crystal system, space group, etc.).

For the given results, we minimized the crystal packing coefficient prediction error using the smoothed L1 loss, defined as $l(x,y) = |x-y| - 0.5$ for $|x-y| > 1$ and $l(x,y) = (x-y)^2/2$ for $|x-y| < 1$. This loss function improves on the standard L1 loss with smoothly converging gradients about zero, while also avoiding the very large losses and accompanying gradients which sometimes destabilize training with the L2 loss. We prioritize stability here, but generally observe comparable performance between the two losses when models converge. The training dataset was generated according to the settings in Table 2, except molecules extracted from crystal entries with nonstandard crystal settings were allowed, resulting in 198,000 samples. The dataset was split 80:20 into training and test datasets. Training is remarkably robust to changes in the model architecture, model/datasets seeds, batch sizes and learning rates, with most runs producing very similar metrics. We always attempt to overfit the model and save the checkpoint with the lowest test loss for evaluation. Training details of the presented runs are given in the Supporting Information.

We chose to model the crystal packing coefficient $c_{pack} = \frac{V_{mol} \cdot Z}{V_{cell}}$ instead of the cell volume or density because these are trivial functions of $c_{pack}$; $V_{cell} = \frac{V_{mol} \cdot Z}{c_{pack}}$, $\rho_{cell} = \frac{m_{mol}}{V_{mol}} \cdot c_{pack}$. Presenting these in lieu of the packing coefficient can obscure the real performance of the model.

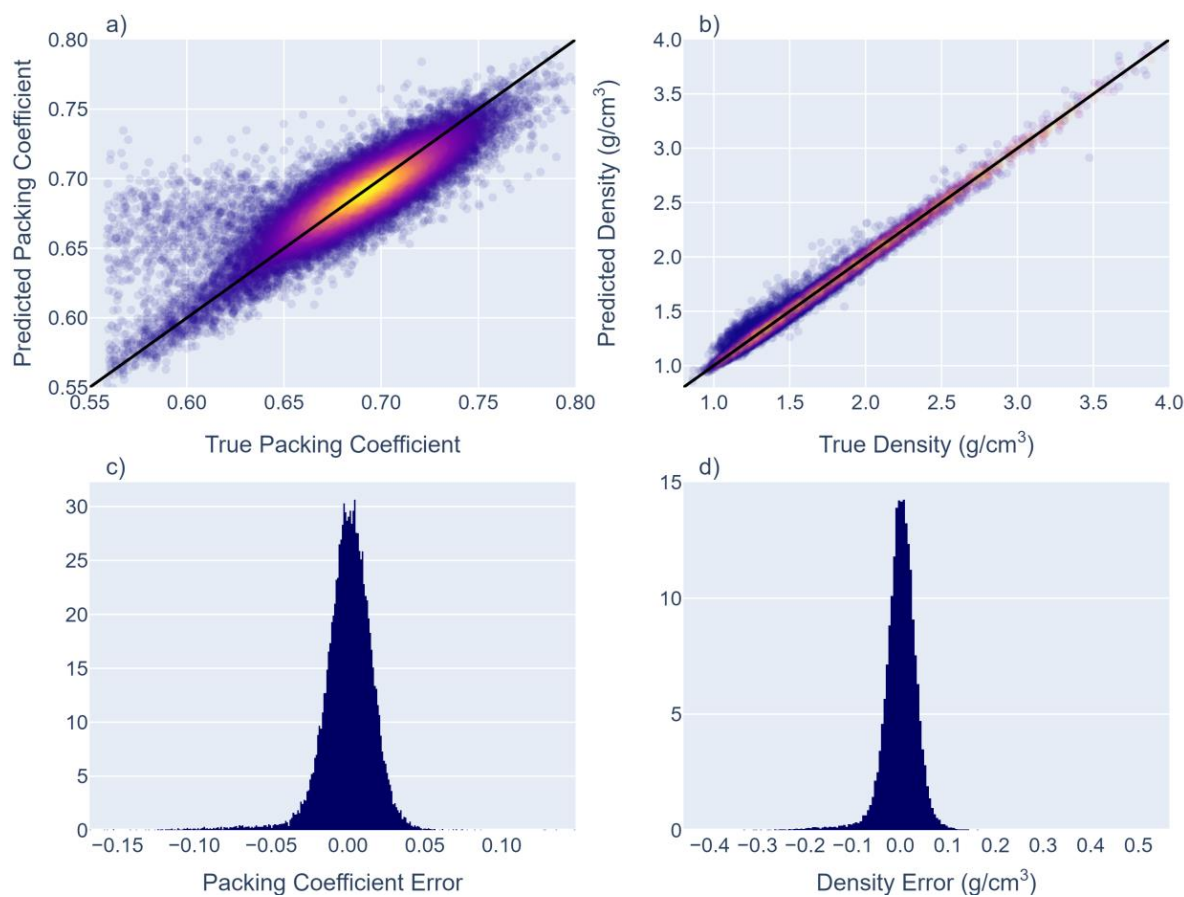

Figure 3: Prediction traces and error distributions from the test dataset (37k samples) of the packing coefficient (unitless, left) and density (g/mL, right). Black diagonal lines correspond to a perfect fit, $R = 1, \text{slope} = 1, \text{predictions} = \text{targets}$.

Figure 3 and Table 3 summarize our results on packing coefficient modelling. Note the apparent improvement going from the packing coefficient to the raw density, despite identical underlying physical information.

Table 3: Summary of regression fit

| Metric | $c_{pack}$ | Density |
|---|---|---|
| MAE | 1.74% | 1.74% |

| MAE σ | 0.0191 | 0.0191 |
|---|---|---|
| Regression R | 0.853 | 0.992 |
| Regression Slope | 0.727 | 0.986 |

Prior works[3,29,30] have already undertaken the modelling of crystal density / unit cell volume from molecular information, with results of generally comparable accuracy. These studies were, however, conducted and validated using much smaller datasets or datasets with only a few types of molecules. Furthermore, the models incorporated only molecule-level features hand-selected by the researchers, that is, they are given a molecule representation rather than learning one directly from the geometry, as in a DGNN.

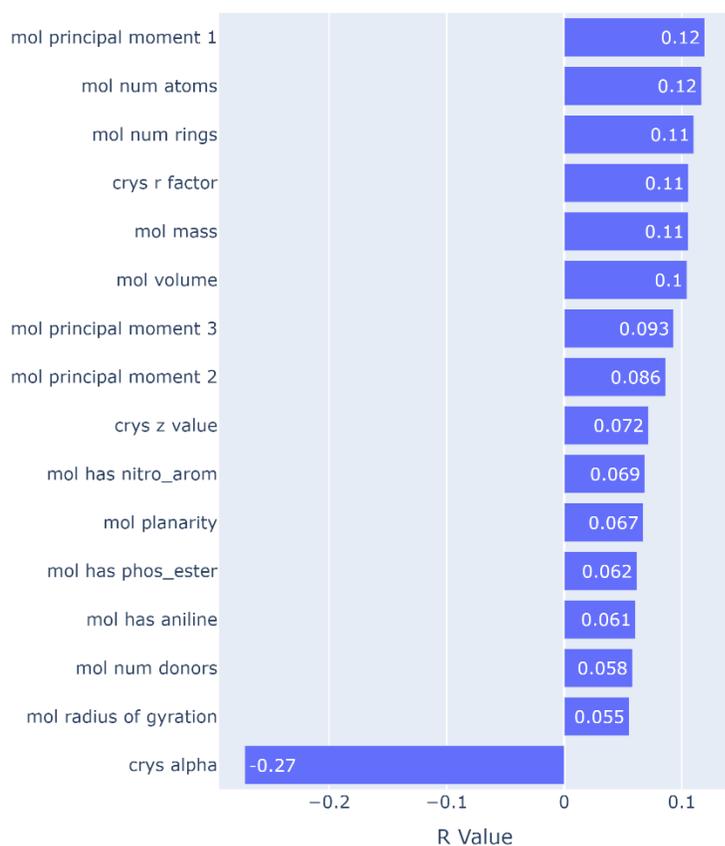

Figure 4: Pearson correlations of losses with sample features, omitting those with absolute values less than 0.05, or features with less than 5% incidence in the test dataset.

In Figure 4, we show the Pearson correlations between per-sample losses and a list of molecule and crystal-level features with non-trivial incidence and correlations, including elemental composition and functional group incidence. We mostly observe weak correlations between the prediction performance and such features, supporting the evidence from Figure 3 that the model is able to generalize well to a wide variety of molecules. There is a very weak positive correlation (~0.1) between error and several factors related to molecule size, including the magnitude of principal moments, number of atoms, and number of rings. There is a stronger negative correlation with the packing coefficient itself, indicating that the model has superior performance on denser crystals. This again corroborates Figure 3, where we see a longer tail of errors for more diffuse crystals.

While we omitted a thorough hyperparameter search, the models demonstrate remarkable robustness across several dimensions. Notably, after cutting the atom and molecule input features down to only the coordinates, atomic number, and molecule volume, accuracy falls by less than 5%. This is quite promising for practical deployment of such a model in a practical search, as depending on user preferences a detailed featurization can be much more expensive than evaluating the DGNN itself. The results are similarly insensitive to changes in the number of graph convolution layers and to increases in the dimension of the model. This consistently high performance calls back to the question, along the lines discussed in Ref. [29] on whether we are perhaps approaching the fundamental limit of accuracy for this type of modelling with the available datasets.

Our density estimation model is fast, accurate, and general, and could be used to support a CSP pipeline by radically tightening the range of likely densities over which to search. Notable absences in our modelling are temperature and pressure, which obviously have an influence on packing density. We discuss in detail the reason for their omission in Appendix A, primarily focusing on inconsistency in the dataset. The high accuracy and particularly the low error variance of our model suggests that their overall effect is significantly less important than the molecular structures themselves.

A further option is conditioning the density prediction on a particular choice of space group to answer the question "what is the predicted density of molecule *f* in space group $P2_12_12_1$?". While technically straightforward, the lack of available data for many space groups may limit the generalization performance of such a model.

## 5. Crystal Scoring

For the scoring and ranking of proposed molecular crystals, we use MolXtalNet-S, incorporating the molecular crystal graph convolution elucidated in Section 2. We further use the dataset generated via the conditions in Table 2, resulting in a training dataset with ~130k samples and a test set with ~30k samples. An extra validation set of ~32k samples is constructed with the same filters using the submissions from Blind Tests 5 and 6.

Training is undertaken in equal sized batches of "real" and "fake" 3x3x3 supercells, with real samples taken from the CSD and fake samples generated by one of our two crystal generators, Gaussian or distorted crystal, between which we alternate with 50% probability. Training runs until overfit or test loss saturation, and we select the checkpoint with the best test loss for evaluation. More details of training methods are given in the Supporting Information.

MolXtalNet-S outputs two raw values indicating on the probability that a given crystal is a real experimental sample or a fake synthesized by our generators. These are normalized by the softmax function,

$$P(y_i) = \frac{e^{y_i}}{\sum_j e^{y_j}},$$

and a loss is computed via the cross-entropy function. The softmax function returns a value between 0 and 1 which is difficult to visualize since, for a well-trained discriminator, almost all values are clustered very close to 0 or 1. To aid in visual discernment, we stretch out the values near 0 and 1 with a function on top of the softmax output, returning a more readable score,

$$S_0(\mathbf{y}) = \tan[(P(y_1) - 0.5) \cdot \pi], \tag{13}$$

$$S(\mathbf{y}) = \text{sign}(S_0(y_1)) \cdot \log_{10}(1 + \text{abs}(S_0(y_1))). \tag{14}$$

This function saturates to 64-bit numerical precision near $\pm 16$, with a softmax value of 0.5, indicating a 50-50 chance of a sample being real or fake, sitting at 0.

A simple method to evaluate crystal quality is to check for overlap in intermolecular vdW radii, as described in the introduction. We undertake such an analysis as a check against our model outputs, with the overall per-crystal score computed as

$$S_{\text{vdW}} = -\log\left[\frac{1}{N_{pairs}} \sum_n^{N_{pairs}} \max\left(0, e^{-\frac{(r_n - r_n^{vdW})}{r_0}} - 1\right)\right], \tag{15}$$

with $r_n$ as intermolecular atom pair distances within a 6Å range, $r_n^{vdW}$ as the sum of van der Waals radii in between atoms in pair $n$, supplied via RDKit, and $r_0$ a scaling factor here set to 1Å. This score saturates to $+\infty$ if all intermolecular ranges are equal or longer than the vdW radii. We therefore apply a clip near 14 to assist visualization.

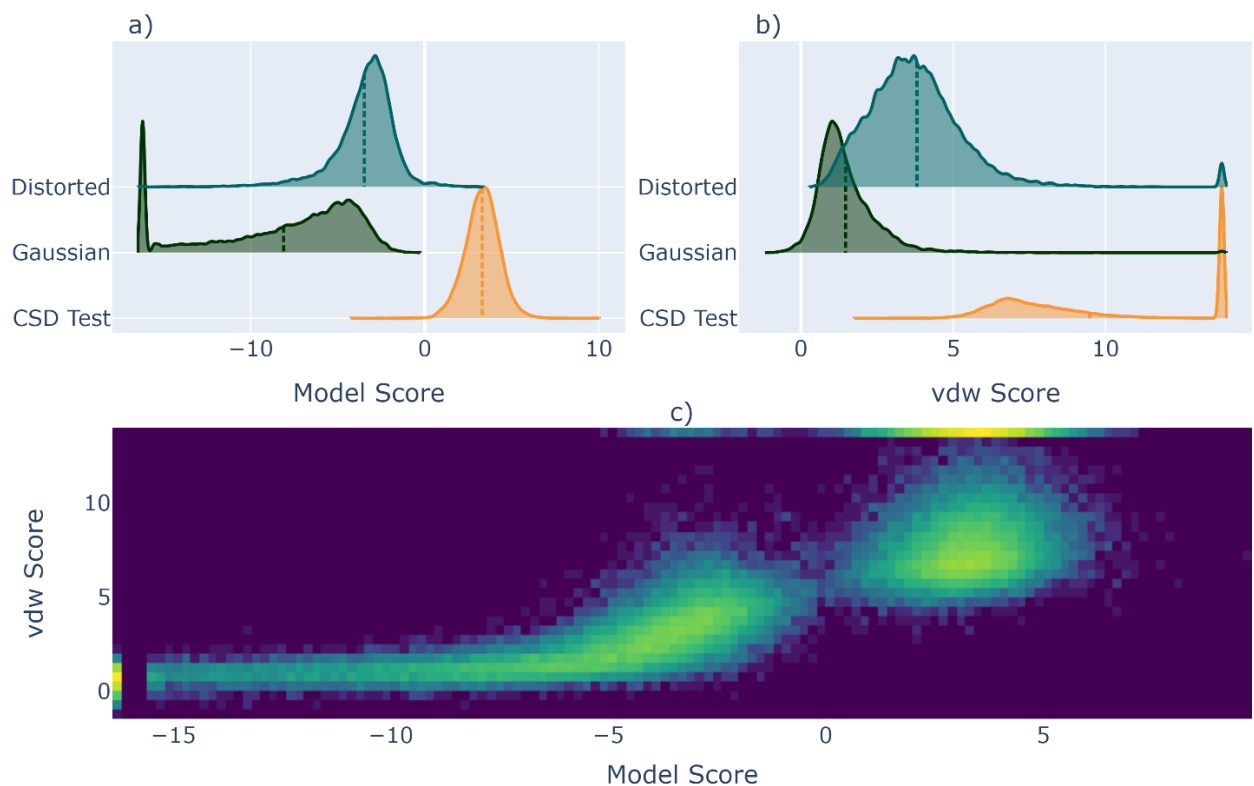

Figure 5: Distribution of model and vdW scores for the CSD test dataset (Real) and fake test sets (Distorted and Gaussian) in a) and b) respectively, with the vdW scores clipped to a maximum near 14 in b). c) shows the 2D distribution for both real and fake test sets. Vertical dotted lines are the distribution means.

In Figure 5, we show typical results on the test datasets of real and fake samples for a well-trained MolXtalNet-S, with the crystal distortion both in training and evaluation set to $c_{dis} = 0.1$. The distributions presented in panel a) indicate that the model decisively rejects almost all of the "fake" samples generated by the Gaussian model. The distorted crystals are rated higher than the Gaussian ones on average, as the applied distortion $c_{\text{dis}}$, is only moderately strong. Since these samples are generated from ostensibly high-quality originals, those that are only subtly distorted should be near but not on the experimental optima. Such samples should teach the model better discernment than the Gaussian samples, whose molecular orientations are practically random and, therefore, almost never respect vdW radii. Another notable trend is the positive correlation between the vdW score and the model score at intermediate values of vdW score, which shows that the model punishes crystals depending on their degree of vdW overlap. At vdW scores above ≈ 5, where atoms largely respect vdW radii, this correlation disappears, as the vdW score loses resolving power. The model on the other hand retains discrimination capability even within samples which respect vdW volumes.

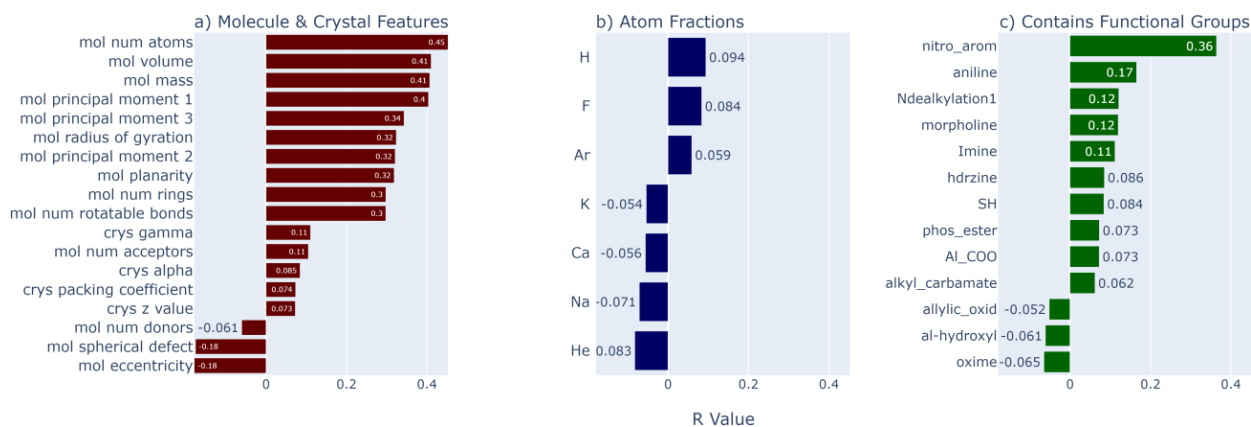

Figure 6: Pearson correlations of discriminator scores with various sample features, omitting those with absolute values less than 0.05, or features with less than 5% incidence in the test dataset. All functional groups within RDKit Fragments module were tested.

In Figure 6, we scan over molecule and crystal features within the test set of real CSD crystals to see what may be favored by MolXtalNet-S. Besides phenyl rings, the model has only weak preferences for specific chemical elements or functional groups, a highly desirable trait in a general-purpose model. This conclusion is supported by a more detailed breakdown in the Supporting Information.

MolXtalNet-S does have a clear preference for flatter and especially larger molecules. There are significant correlations between the model score and the number of atoms per molecule, molecule volume, principal moments, and mass, which are all mutually correlated and related to overall size. This makes comparison between crystals composed of different molecules based on model score more difficult unless one explicitly standardizes by molecule size. We hypothesize this preference arises from the nature of the model training. The DGNN functions by exchanging information in the neighborhood of each atom and reading out this information via global aggregation. Since both real and fake training samples are composed of rigid molecules in realistic conformations, intramolecular correlations will be largely satisfied in every atom in every sample. Put simply, all intramolecular interactions will be rated as "good", with realistic bond lengths and angles, with only intermolecular interactions having the possibility of being "good" or "bad". In larger molecules, there are proportionately fewer atoms exposed to intermolecular vs intramolecular interactions within the convolution cutoff radius, $r_c$. Therefore, there are proportionately more atoms that automatically read as "good" under the model, even in a genuinely low-quality sample.

This size favoritism exists within the distribution of real crystals which are already highly scored and well separated from the fake examples and, therefore, this effect will only be relevant between structures which are already reasonable.

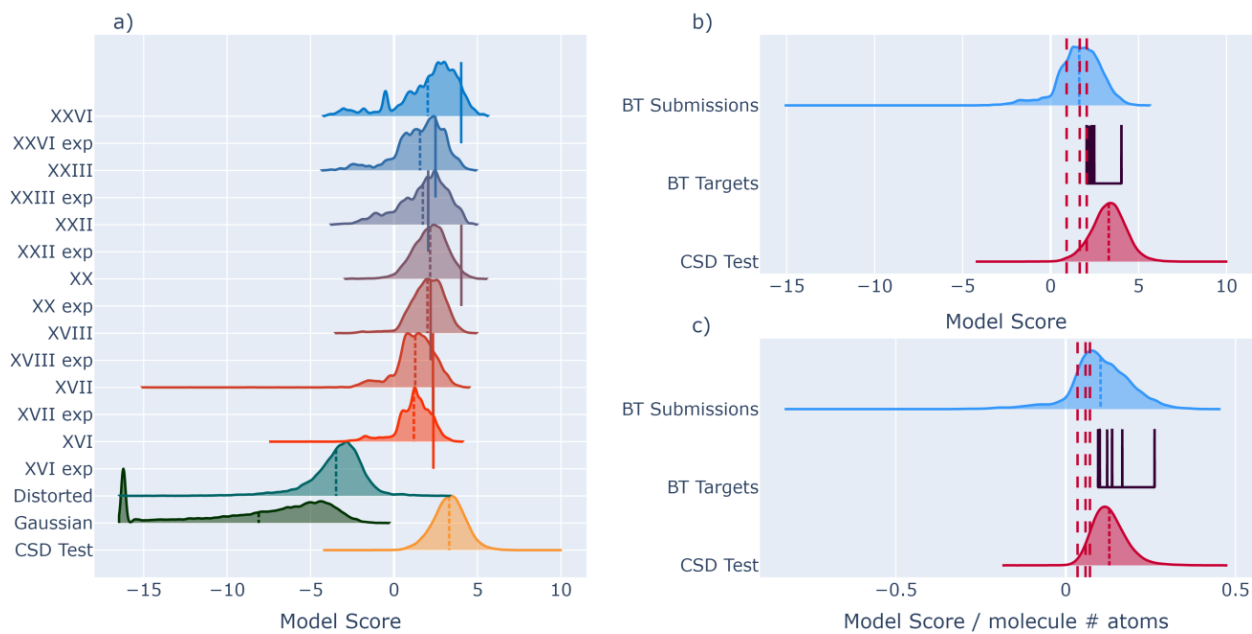

Figure 7: Blind Test submissions scores distributions and means (dashed lines) as well as targets (solid lines). All targets and test data in panel a), combined distributions of test data and all submissions in panels b) and c), with the scores in c) normalized by molecule size. Red dashed lines in b) and c) correspond to the 1%, 5% and 10% quantiles of the CSD test distribution.

In Figure 7, we show the distributions of scores for the $Z' = 1$ Blind Test submissions from Blind Tests 5 and 6. Chemical diagrams for the relevant targets are given in Appendix D. The experimental targets are not always at the top of the distribution of their respective submissions, indicating that the model has limited discrimination capacity at the very-high-end. However, large fractions (77% on average) of all submissions are scored below their respective targets, highlighting that even among a batch of crystals with reasonable densities and respect for vdW radii, the experimental structure is favored by the model. This, combined with the low cost of evaluation, makes MolXtalNet-S an attractive support tool to energy evaluations in a structure search.

The distribution of scores for Blind Test submissions diverges notably from the CSD samples and this separation can be widened by overfitting the model on the training data. This, however, comes at the cost of seepage of CSD crystals into the low-quality regime, which we avoid for now to maintain maximum model robustness, and because it is not necessarily clear to what degree they should diverge.

Excluding submissions below the 5% quantile of the CSD test distribution would efficiently filter a large proportion of proposed crystals, 48% in this case. This is, however, partly illusory, driven by the different relative means of the distributions for each target, due to the model's preference for larger molecules. Normalizing the scores by the number of atoms per molecule tightens the spacing between the submissions and CSD test distributions and reduces variance between the BT targets, while also reducing the fraction of submissions below the 5% CSD test quantile to 27%, which is still substantial.

A graph model could, therefore, be used as a filter, similarly to a vdW check, at only slightly greater cost. The major difference is in physical nuance: a vdW check is a quite crude tool that cannot distinguish between crystals such as the Blind Test submissions which already respect atomic vdW radii. MolXtalNet-S, on the other hand, maintains remarkable resolving power even between samples that are already of generally good quality, constituting the work of many researchers and millions of processor hours, even effectively filtering one third of them at the 5% quantile threshold. Beyond the search for the "most stable" experimental structure, a DGNN model could certainly be used in the search for the stable polymorphs of a given molecular crystal, in the pre-screening, search and even refinement stages, where its speed and generality would be an asset.

MolXtalNet-S loses some discrimination sensitivity once within the bulk of the CSD scores distribution. It is not clear whether this is a limitation of the model or training protocol, whether we require even more sophisticated "fakes" to discriminate against, or whether there is a quality issue within the experimental data. In principal, a deep GNN should be able to learn very complex functions of molecular geometry[42], though details of adequate training are far from trivial, and the field is evolving rapidly even now[23].

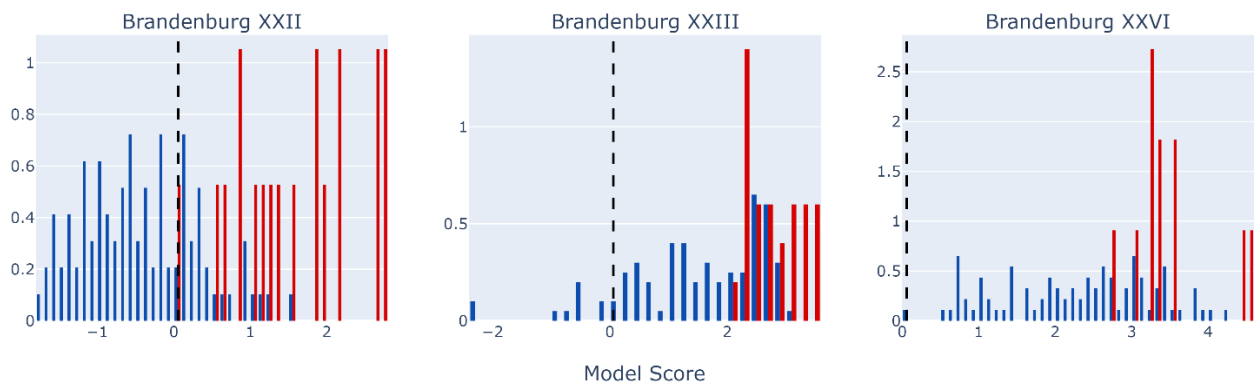

Figure 8: Model score distributions for particular CSD Blind Test 6 submissions. Blue bars are the first submission and red the second. Black dashed line is the non-normalized 5% quantile score for the CSD test dataset.

For Blind Test 6, groups were allowed to submit multiple rankings, potentially with different methodologies. This difference in methodology is particularly pronounced in the response of the models to the Brandenburg submissions for targets XXII, XXIII and XXVI, shown in in Figure 8. The starting structures were taken from the Price submission, and reoptimized/reranked at two levels of theory. The second submission in red, with the higher level of theory, was scored significantly higher by the model. The fact that the model discriminates between submissions optimized at different levels of theory is encouraging as, assuming that the higher level of theory gives better agreement with experiment, we see this correctly captured by the model. Scores vs. rankings for all the $Z' = 1$ submissions to Blind Test 6 are shown in the Supporting Information.

These results show the potential of DGNNs as part of a CSP pipeline. If this tool had been used in the Blind Test 5 and 6 submissions, it would have filtered out a meaningful fraction of the submitted structures, depending on applied tolerances. The group-specific analyses suggest additional practical uses for this type of discriminator model. First is a check on CSP methods in general. For example, when testing a new force-field, one could score some of its optimized structures via the DGNN. If the distribution of scores is far from that for experimental structures, this is a clue that the force-field is mis-fit and perhaps setting potential minima incorrectly. A second use for such models is directly in a search tool, supplementing or replacing energy evaluations which are currently used. For example, in a straightforward Markov chain Monte Carlo search we search for the global optimum of some score function, generally the potential energy, by a chain of discrete jumps in the space of crystal parameters, $C$. There is no reason why the potential energy could not be replaced with a stability score from a DGNN, and indeed we have done initial tests in this direction. These models are inexpensive to run on a modern GPU, with throughput naively on the order of hundreds of thousands of structures per hour.

## 6. Conclusions

In this contribution, we introduced and applied geometric deep learning methods to the study of molecular crystals. Such methods combine speed, quality, and wide applicability, making them powerful tools for the acceleration of molecular crystal structure prediction. Currently, state-of-the-art CSP methods can be extremely expensive, expending several months and millions of CPU hours on single crystals. Our approach is therefore timely, offering inexpensive yet powerful tools for the support and acceleration of CSP pipelines. Further, both MolXtalNet-D and MolXtalNet-S demonstrate remarkable equanimity and generalization performance to practically all common functional groups and atom types in the CSD.

The analyses in Sections 4 and 5 demonstrate the general capabilities of DGNNs for molecular crystal property prediction from single molecules and crystal configurations. MolXtalNet-D predicts the packing density of bulk molecular crystals from only single conformer information and achieves state-of-the-art performance with minimal tuning. The new molecular crystal periodic convolution allows us also to model molecular crystal graphs, and we trained MolXtalNet-S, an efficient model for scoring molecular crystals. As the most expensive step in a DGNN forward pass is generally the construction of the initial radial graph, our trained models are only marginally more expensive than simple volume exclusion methods, such as the vdW radius check, while incorporating significantly more physical nuance, and retaining discrimination sensitivity even between already reasonable crystal samples. As discussed in Section 5, performance appears limited by the quality of available training data, and there remains significant room for algorithmic improvement and application of additional computational power.

Data processing, training, and analysis scripts are all available in our [GitHub repository](#). The training data can be sourced via the CSD Python API, which is for now a necessary package in our processing pipeline. Beyond density prediction and sample ranking, our code provides necessary tools for general molecular crystal learning tasks, which we plan to leverage in later studies. Particularly useful is our unit cell builder. Written and optimized in PyTorch, this module is fast, parallel, and differentiable, and therefore suitable for within-loop crystal generation and backpropagation tasks. The state-of-the art of GNN architectures is currently evolving very quickly and may soon enable further performance gains. Extensions of our approach to $Z' \neq 1$ structures and disordered crystals are both possible with relatively straightforward modifications to the existing algorithm. A key next step will be the development of more advanced generative models for molecular crystal configurations, both to accelerate CSP by extremely rapid sampling of high-quality initial candidate structures, and as a proof of concept for condensed phase molecular materials generation.

**Appendix A: Molecule graph featurization**

A list of atom and molecule-wise features included in our modelling is provided below. In general, features which are floating point or integers are standardized before training, whereas Booleans are left as-is.

Atom-wise

- Atomic number
- Mass
- Atom is H-bond acceptor
- Atom is H-bond donor
- Valence
- vdW radius
- Atom is on a ring
- Electronegativity

Molecule-wise

- Volume
- Mass
- # Atoms
- Molecule is chiral
- # Rings
- # H-bond donors
- # H-bond acceptors
- # Rotatable bonds
- Planarity
- Polarity
- Asphericity
- Eccentricity
- Radius of gyration
- Principal moments 1-3

A notable absence in our features list is the experimental temperature and pressure from the CSD. While less than ideal, it does not appear at this time that there is a straightforward and consistent way to include them. A large fraction of CSD entries lack one or both of these datapoints, which in any case are generally under-specified. That is, it is not clear at all in general how one should connect the recorded crystal temperature and pressure with the actual crystallization & measurement process. Therefore, despite their clear relevance to the properties of crystals, we omit them from our analysis for now, and accept the necessary loss of precision of the final models.

**Appendix B: Fast and differentiable supercell builder**

Likely the most important piece of our pipeline for future work, the cell builder allows us to propose and build cells in a fast and differentiable way, so that one can train directly on crystal generation tasks.

The first stage is the optional cleaning of incoming cell parameters, $C$, defined in Section 2. This may be necessary if the parameters are being generated by some automated protocol, as in our case, which may not adhere to all physical limitations such as positive fractional positions, reasonable angles, and requirements of the crystal system. We therefore "clean" the parameters in the following ways: $(a, b, c)$ are enforced to be positive via softplus function, $(\alpha, \beta, \gamma)$ are enforced to be between 0 and $\pi$ via a hardtanh function, likewise the molecule orientation parameters $(\phi, \psi, \theta)$ between 0 and $2\pi$, and the molecule centroid position $(\bar{x}, \bar{y}, \bar{z})$ is fixed between 0 and 0.5 in fractional coordinates by another hardtanh. Further, the cell lengths and angles are fixed according to the pre-selected crystal system of the sample, e.g., a cubic crystal will have cell lengths enforced equal and angles at $\frac{\pi}{2}$.

Next, a single molecule is placed in the unit cell with shape defined by $(a, b, c, \alpha, \beta, \gamma)$, at the position and orientation $(\bar{x}, \bar{y}, \bar{z}, \phi, \psi, \theta)$ as described in Section 2. The general Wyckoff symmetry is used to pattern the unit cell up to $Z$ molecules, and finally the unit cell is patterned out to the *N*x*N*x*N* (usually 3x3x3) supercell for discriminator scoring and training.

While not strictly necessary for this study, which could have been undertaken via a pre-built dataset, this tool is fast and parallel, allowing for straightforward on-line usage inside a training or evaluation loop. Further, it is written and tested in PyTorch, and gradients may pass between a model which generates the cell parameters to a loss function applied on the resulting crystal, opening the door for the training of generative models for molecular crystals.

**Appendix C: Procedure for polymorph filtering**

Polymorphs of the same molecule are identified by scanning CSD identifiers for duplicate prefixes (leading six letters of the identifier). CSD identifiers are formatted as "XXXXXX##", with polymorphs, if any, each assigned a number with a shared prefix. To select a single representative polymorph, we select the structure with the best r-factor, which shares the same space group as the oldest submitted structure, the idea being the first-discovered structure is likely the most stable. Polymorph filtering in this way is useful to avoid bleed-over between training and test datasets, though in practice our models generalize rather well without overfitting, so it tends to make little empirical difference.

**Supporting Information.** Attached PDF contains supporting information including details of model training, target and group-wise ranking analysis for the Blind Test 6 targets, and analysis of model score vs a wide range of functional groups and elemental compositions.

**Appendix D: Chemical structures of single component Blind Test 5 and 6 targets**

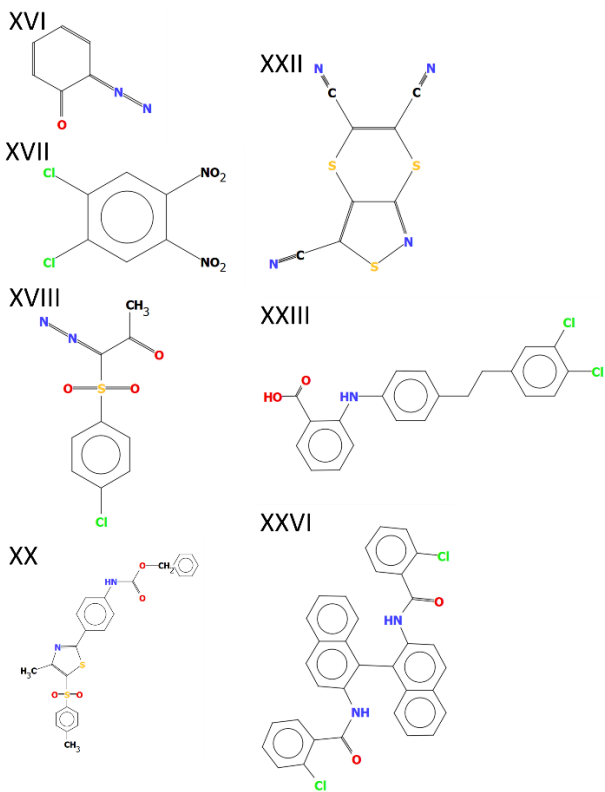

Figure 9: Chemical diagrams for the Blind Test 5 and 6 structures used in model evaluation. Roman numerals indicate CSD Blind Test label.

AUTHOR INFORMATION

**Corresponding Author**

Correspondence should be addressed to mark.tuckerman@nyu.edu.

**Author Contributions**

The manuscript was written through contributions of all authors. All authors have given approval to the final version of the manuscript.

**Funding Sources**

Work of MK funded by a Natural Science and Engineering Research Council of Canada (NSERC) postdoctoral fellowship. JR acknowledges financial support from the Deutsche Forschungsgemeinschaft (DFG) through the Heisenberg Programme project 428315600. JR and MET acknowledge funding from grants from the National Science Foundation, DMR-2118890, and MET from CHE-1955381.